\documentclass[reprint,amsmath,amssymb,aps,nofootinbib]{revtex4-2}
\usepackage{graphicx}
\graphicspath{ {images/} }
\usepackage{epstopdf}
\usepackage{hyperref}
\usepackage{multirow}

\usepackage{color}
\definecolor{green}{rgb}{0,0.5,0}
\usepackage{xcolor}
\usepackage{textcomp}

\begin{document}

\title{Anomalous diffusion in polydisperse granular gases: Monte Carlo simulations}
\author{Anna S. Bodrova$^1$ and Alexander I. Osinsky$^2$}
\date{\today}

\address{$^1$ Moscow Institute of Electronics and Mathematics, HSE University, 123458, Moscow, Russia}

\address{$^2$ Skolkovo Institute of Science and Technology, 121205, Moscow, Russia}

\begin{abstract}
We investigate both ensemble and time-averaged mean-squared displacements of particles in a polydisperse granular system in a homogeneous cooling state. The system contains an arbitrary number of species of different sizes and masses. The collisions between granular particles are described in terms of the models both of constant and time-dependent restitution coefficients. In our study, we use a powerful low-rank
algorithm that allows for efficient simulation of highly polydisperse granular systems. The Monte Carlo simulations are in good agreement with the analytical results.
\end{abstract}

\maketitle

\section{Introduction}

There are numerous examples of granular materials \cite{GranRev, PhysGranMed, SanPowGr, DryGranMed} appearing in nature and used in various technologies, such as sand and stones in the building industry, rice, sugar, salt, and coffee in the food industry, and different kinds of powders in chemical and cosmetic production. The surfaces of Mars \cite{mars}, other planets and satellites are covered by granular dust. 

Granular gases represent diluted granular systems \cite{book}, where the distance between their components significantly exceeds their size. The total packing fraction, $\phi<0.2$, is the ratio of the total volume of a set of objects packed into a space to the volume of that space. 

Initial studies on granular gases were devoted to one-component granular gases owing to their simplicity \cite{book}. 
However, in nature and technology, granular systems are mostly polydisperse and consist of various particles with different sizes and masses. In systems such as large interstellar dust clouds \cite{inter}, protoplanetary discs, and planetary rings \cite{ringbook,rings, pnas}, populations of asteroids may be considered as granular gases \cite{aster}. 

Collisions of granular particles are dissipative, which leads to a decrease in the mean kinetic energy of granular systems, usually termed granular temperature. Therefore, the granular gas cools down. In the later stages of evolution, clusters and vortexes may form in the system; however, in the initial state, the system remains homogeneous. This initial state of evolution is termed as the homogeneous cooling state.

The theory of granular gases was developed as an extension of ideal gas models, in which dissipation during interparticle collisions was considered. Thus, granular gas in a homogeneous cooling state represents a fundamental physical system in statistical mechanics and can be regarded as a reference model in granular matter physics \cite{mehta}.

Despite the theoretical significance of studying granular gases, they are relatively difficult to experimentally obtain. Granular gases can be investigated by placing granular matter in containers with vibrating \cite{vib1, vib2} or rotating \cite{rot} walls and applying electrostatic \cite{el} or magnetic forces \cite{magn1, magn2}. To obtain force-free gases, they are placed in a microgravity environment in drop towers \cite{drop1, drop2} on sounding rockets \cite{Sperl, rocket1, rocket2, rocket3}, parabolic flights \cite{flights1, flights2, flights3, flights4, flights5, flights6}, and satellites \cite{sat}. However, such experiments are expensive and difficult to implement. In current studies, the microgravity environment cannot persist long enough and the obtained trajectories of the particles are relatively short. Hopefully, the development of corresponding technologies will lead to higher quality data in the future. 

Modern computer algorithms allow us to investigate the behavior of granular media over longer periods. Typically, molecular dynamics, event-driven algorithms, or direct simulation Monte Carlo (DSMC) methods can be used to simulate granular systems \cite{compbook}. In this study, we focused on DSMC simulations \cite{Bird} using a low-rank technique. It was first applied to the solution of Smoluchowski differential equations \cite{rank1, rank2, rank3}, and was recently modified for Monte Carlo simulations of aggregation \cite{mediag}. Here, we use the same idea for the DSMC simulations of diffusion in granular mixtures.

Let $\textbf{R}(t)=\int_0^t\textbf{v}(t^{\prime})dt^{\prime}$ be the displacement of the particle and $\textbf{v}(t)$ be its velocity. Owing to dissipative collisions, the motion of granular particles is anomalous with a non linear dependence of the mean-squared displacement (MSD) at time $t$ \cite{ralf, sok, eli, georges, franosh}:
\begin{equation}\left\langle R^2(t) \right\rangle \sim t^{\alpha}\,,\qquad\qquad \alpha\ne 1\end{equation}
The ultraslow motion occurs with a logarithmic time-dependence \cite{ido,ultraslow, jeon}:
\begin{equation}\left\langle R^2(t) \right\rangle \sim \log t\,.
\end{equation}
The motion of particles in a force-free cooling unicomponent granular gas may be either ultraslow or subdiffusive, with $0<\alpha<1$ \cite{annapccp}. The diffusion of granular intruders in binary granular mixtures \cite{garzointruder, annaprl,garzoreview} and granular suspensions \cite{garzointruder23} has been previously investigated. The expression for MSD obtained in a binary mixture \cite{garzointruder} and in a polydisperse mixture \cite{garzointruder24} is valid only in the long-time limit. Considering that current experimental investigations only allow observation of the behavior of cooling granular gas at short times, the applicability of this expression is quite limited. In \cite{anna2024} the MSD was calculated for polydisperse granular mixtures in terms of numerical integration. 

In the current investigation, we provide the full analytical expression for the MSD valid at all time scales and confirm its validity in terms of DSMC simulations, which is the key result of this study. We also derive the time-averaged MSD for granular intruders.

We proceed as follows. In Section II, we review previous studies and provide new analytical results for both MSD and time-averaged MSD in granular mixtures. In Section III, we describe our computer algorithm and discuss its advantages over other simulation methods for highly polydisperse granular systems. Finally, we present our conclusions in Section IV.

\section{Theory}

\subsection{Mean squared displacement}

We assume that the granular particles could be considered as hard spheres. The diameter of the particles is $\sigma_k$. If the particles are made of the same material, their diameters are $\sigma_k=\sigma_1k^{1/3}$. The number densities of species are $n_k=N_k/V$, where $N_k$ is the number of particles of species $k$, $V$ is the volume of the system. The total number density is $n=\sum_k n_k$. In a mixture of $N$ species, the total MSD can be expressed by averaging over partial MSDs:
\begin{equation}\label{MSD}
\left< R^2(t)\right>=\frac{1}{n}\sum_{k=1}^N n_k\left< R^2_k(t)\right>\,,
\end{equation}
where the partial MSD $\left< R^2_k(t)\right>$ take the form \cite{anna2024}
\begin{equation}\label{R2}
\left< R^2_k(t)\right>=6\int_0^tdt_1 D_k(t_1)\left[1-\exp \left(-\frac{\tau_k(t)-\tau_k(t_1)}{\hat{\tau}_{v,k}(t_1)}\right)\right]\,.
\end{equation}

Here the reduced velocity correlation time is
\begin{equation}\label{hattau}
\hat{\tau}_{v,k}(t)=\tau_{v,k}(t)\sqrt{\frac{T_k(t)}{T_k(0)}}\tau_c^{-1}(0)
\end{equation}
and the reduced time $\tau_k$ is introduced according to
\begin{equation}\label{timeres}
d\tau_k=dt\sqrt{T_k(t)/T_k(0)}\tau_c^{-1}(0)\,.
\end{equation}
The inverse mean collision time is
\begin{equation}\label{tauc}
\tau_c^{-1}(t)=4n_1\sigma_1^2g_2(\sigma)\sqrt{\frac{\pi T_1}{m_1}}\,.
\end{equation}
The partial diffusion coefficient of species $k$ may be calculated according to
\begin{equation}\label{diffk}
D_k(t)=\frac{T_k(t)\tau_{v,k}(t)}{m_k}\,.
\end{equation}
The granular temperature $T_k$ of species $k$ is introduced according to \cite{garzoreview}.
\begin{equation}
\frac32 n_k T_k=\frac{m_k\langle v_k^2\rangle}{2} =\int d {\bf v}_k f_k\left({\bf v}_k, t\right)\frac{m_kv_k^2}{2}\,.
\end{equation}
The velocity distribution function $f\left({\bf v}_k,t\right)$  of species $k$ is assumed to be Maxwellian.

The inverse velocity correlation time is given by the sum
\begin{equation}\label{tausum}
\tau_{v,k}^{-1}(t)=\sum_{i=1}^N \tau_{v,ki}^{-1}(t)\qquad\qquad k=1,...,N.
\end{equation}
The terms in the inverse velocity correlation time (Eq.~\ref{tausum}) take the following values \cite{anna2024}:
\begin{eqnarray}\nonumber
&&\tau_{v,ki}^{-1}(t)=\frac{8\sqrt{2\pi}}{3}n_i\sigma_{ki}^{2}g_{2}(\sigma_{ki})\frac{m_i}{m_i+m_k}\frac{T_km_i+T_im_k}{T_k\left(m_i+m_k\right)}\\&&\times
\left(\frac{T_km_i+T_im_k}{m_im_k}\right)^{1/2}\frac{\left(1+\varepsilon_{ki}\right)^2}{4}\,.
\label{tauvconst}
\end{eqnarray}
Here, $g_2\left(\sigma_{ki}\right)$ is the contact value of the pair correlation function \cite{book}, $\sigma_{ki}=\left(\sigma_k+\sigma_i\right)/2$. 

The restitution coefficient $\varepsilon_{ki}$ accounts for the energy loss in the dissipative collisions \cite{GranRev,book}:
\begin{equation}\label{epsdef}
\varepsilon_{ki} = \left|\frac{\left({\bf v}^{\,\prime}_{ki} \cdot {\bf e}\right)}{\left({\bf v}_{ki} \cdot {\bf
e}\right)}\right| \, .
\end{equation}
Here  ${\bf v}_{ki}={\bf v}_{k}-{\bf
v}_{i}$ and ${\bf v}^{\,\prime}_{ki}={\bf v}_{k}^{\,\prime}-{\bf v}_{i}^{\,\prime}$ are the relative velocities of particles of masses $m_k$ and $m_i$ before and after a  collision, respectively, and ${\bf e}$ is a  unit vector directed along the inter-center vector  at the collision instant. $\varepsilon_{ki}=1$ corresponds to perfectly elastic collisions with conserved energy. $\varepsilon_{ki}=0$ accounts for perfectly inelastic collisions. 
A rare case $\varepsilon_{ki}<0$ may occur during oblique collisions \cite{Saitoh}.
For simplicity, the restitution coefficient is assumed to be constant and equal for all colliding species in most granular gas models \cite{book}: $\varepsilon_{ki}=\varepsilon=\rm const$. We use this simplified assumptions below.
The evolution of granular systems with velocity-dependent restitution coefficient is presented in Appendix A.

\begin{figure}\centerline{\includegraphics[width=0.95\columnwidth]{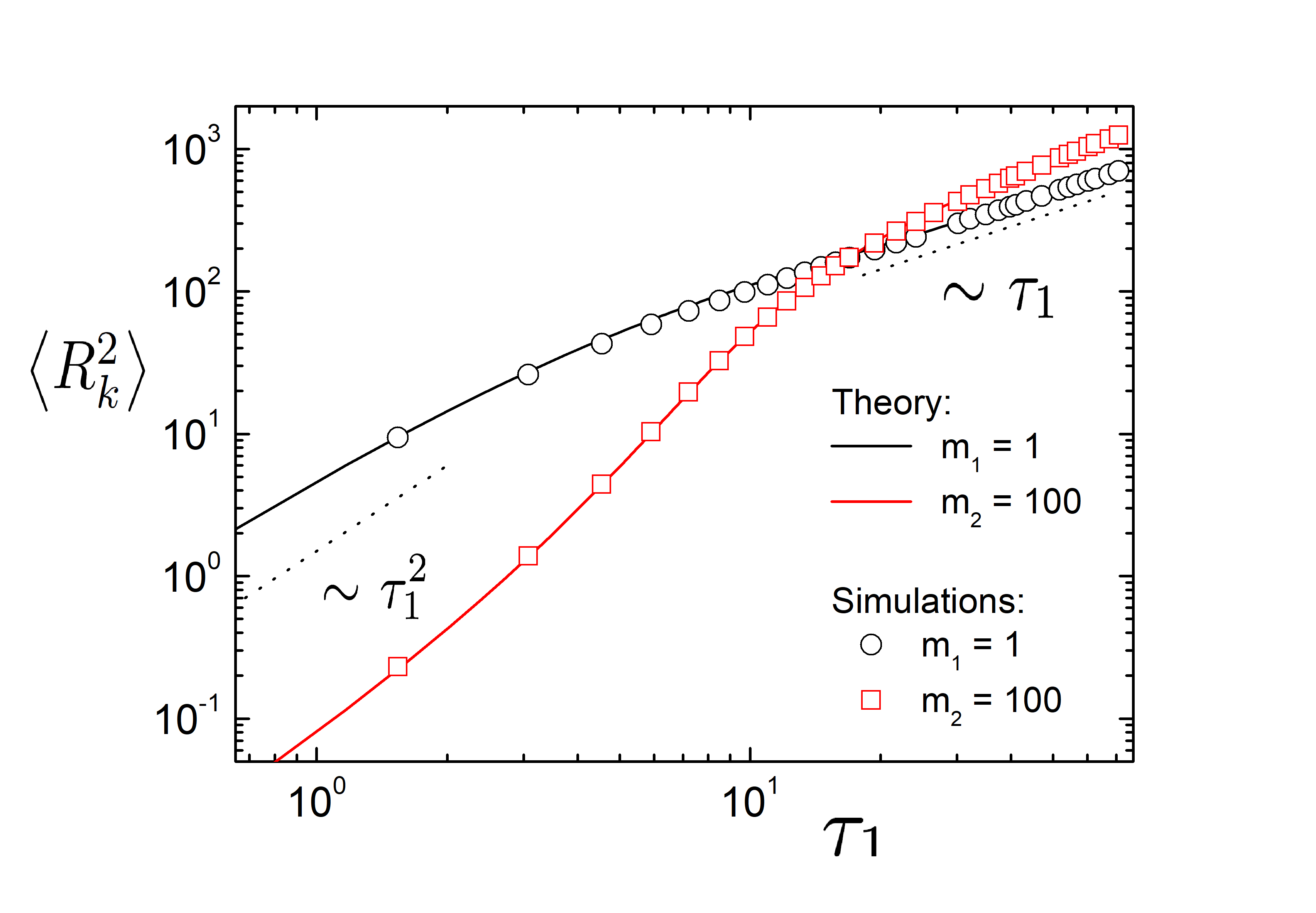}}\caption{Plot of partial MSDs as a function of the rescaled time $\tau_1$, where $d\tau_1=dt\sqrt{T_1(t)/T_1(0)}/\tau_c(0)$. The binary granular mixture of particles, colliding with a constant restitution coefficient, $\varepsilon=0.5$ is considered. The partial number densities of particles are equal: $n_1=n_2=0.1$. The masses of species are $m_1=1$, $m_{2}=100$, the diameters $\sigma_1=\sigma_{2}=1$. At short time the particles move along ballistic trajectories $\left< R^2_k\right>\sim \tau_1^2$, at long times the particles perform normal diffusion $\left< R^2_k\right>\sim \tau_1$ (shown with a dotted line). At the initial time moment, the equipartition holds: $T_1(0)=T_2(0)=1$.
Symbols denote the results of DSMC simulations. } \label{Grescaled}
\end{figure}


\subsection{Granular temperatures}

Owing to dissipative collisions, the granular temperatures of species decrease, whereas equipartition does not hold \cite{book}. The evolution of partial granular temperatures in a mixture occurs according to the following system of differential equations \cite{lev, zippelius, hrenya}:
\begin{eqnarray}\label{sys}
\frac{dT_k}{dt}=-T_k\xi_k\,\qquad\qquad k = 1,...,N 
\end{eqnarray}
The cooling rate is equal to the sum
\begin{equation}\label{eq:sys_xyk}
\xi_k=\sum_{i=1}^N\xi_{ki}\,.
\end{equation}
 The terms $\xi_{ki}$, which quantify the decrease in the granular temperature of species of mass $m_k$ owing to collisions with species of mass $m_i$ are given by the following expression \cite{lev, zippelius, hrenya}:
\begin{eqnarray}\nonumber
&&\xi_{ki}(t) =\frac{8}{3}\sqrt{2\pi}n_i\sigma_{ki}^{2}g_{2}(\sigma_{ki})
\left(\frac{T_km_i+T_im_k}{m_im_k}\right)^{1/2}\!\left(1+ \varepsilon_{ki}\right)\\&&\times
\left(\frac{m_i}{m_i+m_k}\right)\left[1-\frac{1}{2}\left(1+ \varepsilon_{ki}
\right)\frac{T_im_k+T_km_i}{T_k\left(m_i+m_k\right)} \right] \label{xikconst}.
\end{eqnarray}
The cooling rates $\xi_k$ for the constant restitution coefficient become equal after a short relaxation time for all species $k$, leading to a constant ratio of granular temperatures $T_k/T_l$ during the evolution of the system.

First, let us consider the case of initial equipartition when the temperatures of all species are equal at the initial time: $T_k(0)=1$, $k=1..N$. The evolution of the binary mixture, starting with equipartition, is depicted in Fig.~\ref{Grescaled}. In terms of the time scale $\tau_1$ (Eq.~\ref{timeres} with $k=1$) the particles $m_1$ they move ballistically $\sim \tau_1^2$ at short times and diffusively $\sim \tau_1$ at long times. Particles of mass $m_2>m_1$  lose a small amount of energy during collisions with lighter particles. Initially, they have a smaller velocity because of the temperature equipartition; however, as time passes, their granular temperature becomes relatively large, and they start moving with acceleration with respect to the smaller particles. Finally, the temperature ratio becomes constant, and the higher particles undergo normal diffusion, as well as the lighter ones.

\begin{figure}\centerline{\includegraphics[width=0.95\columnwidth]{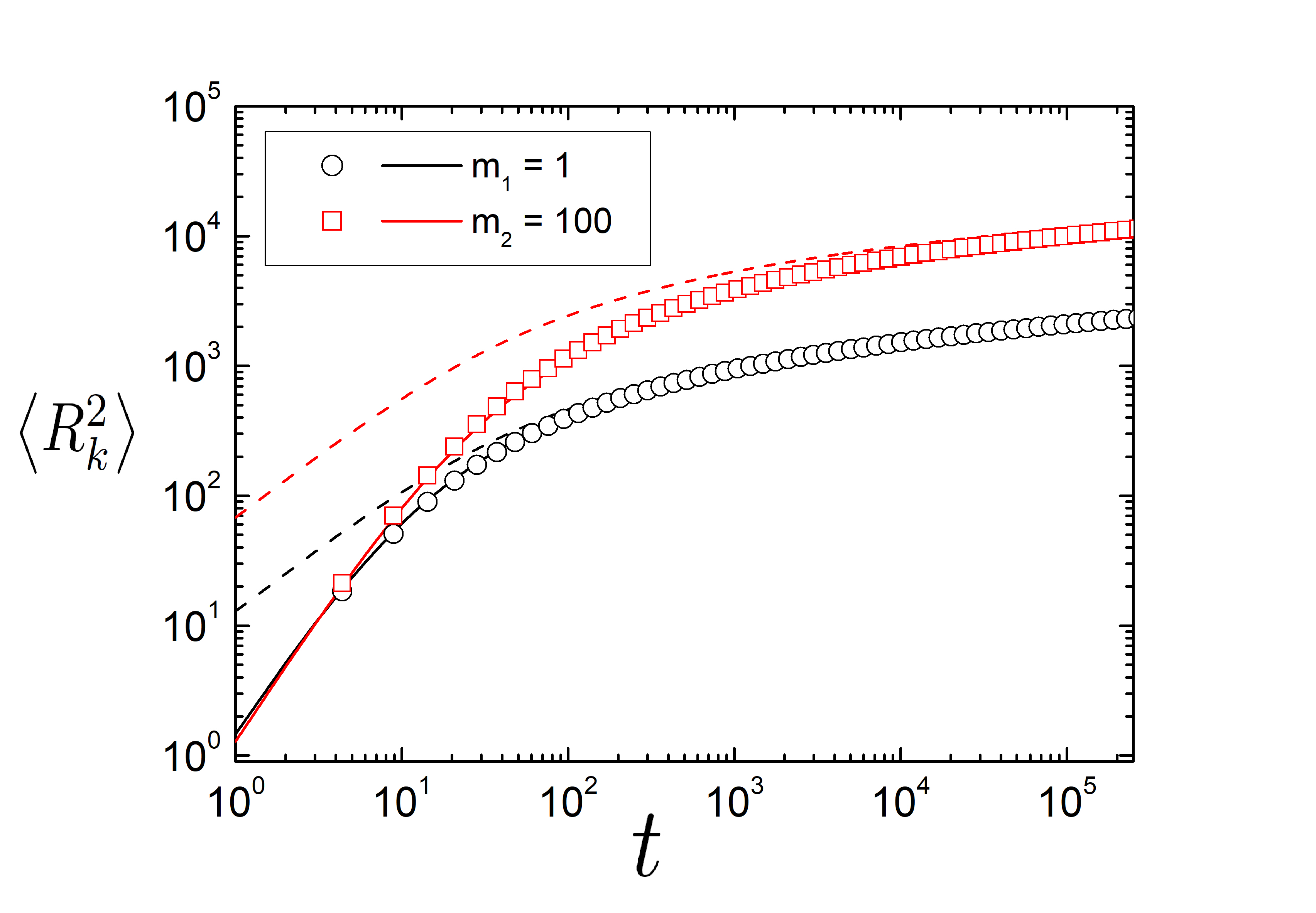}}\caption{Time dependence of partial MSDs in a binary granular mixture of particles, colliding with a constant restitution coefficient $\varepsilon=0.5$. The partial number densities of particles are $n_1=0.1$, $n_2=0.001$ and there are $N_1=10^6$ and $N_2 = 10^4$ particles. The masses of species are $m_1=1$, $m_{2}=100$, the diameters $\sigma_1=1$, $\sigma_{2}=100^{1/3}$. The initial granular temperatures are $T_1(0) = 0.554$, $T_2(0) = 45.58755$. The temperature relaxation time $\tau_0=19.258$. The dashed line corresponds to MSD given by Eq.~(\ref{R2anallong}). Symbols denote the results of DSMC simulations.} \label{Grealnew}\end{figure}

\subsection{Evolution with a constant cooling rate}

Now, let us discuss the case where we start our measurement when the constant ratio of temperatures is already achieved, and the ratio
\begin{equation}\label{gamma}
\gamma_{ik}=\frac{T_i(t)}{T_k(t)}=\frac{T_i(0)}{T_k(0)}
\end{equation}
remains constant throughout the evolution of the system for all species, $k,i=1..N$, presented in the system.
The evolution of partial granular temperatures occurs according to Haff's law \cite{book, haff}:
\begin{equation}\label{TkHaff}
T_k(t)=T_k(0)\left(1+\frac{t}{\tau_0}\right)^{-2}\,.
\end{equation}
By integrating the rescaled time (Eq.~\ref{timeres}), it becomes equal for all species:
\begin{equation}
\tau_k=\frac{\tau_0}{\tau_c(0)}\log\left(1+t/\tau_0\right).
\end{equation}
The cooling rates given by (Eqs.~\ref{eq:sys_xyk}-\ref{xikconst}) attain the form
\begin{eqnarray}
\xi_k(t)=\sum_{i=1}^N\hat{\xi}_{ki}\left(1+\frac{t}{\tau_0}\right)^{-1}=\xi_l(t)\,.
\end{eqnarray}
Here the coefficients
\begin{eqnarray}\nonumber
&&\hat{\xi}_{ki} =\frac{8}{3}\sqrt{2\pi T_k(0)}n_i\sigma_{ki}^{2}g_{2}(\sigma_{ki})
\left(\frac{m_i+\gamma_{ik}m_k}{m_im_k}\right)^{1/2}\!\left(1+ \varepsilon_{ki}\right)\\&&\times
\left(\frac{m_i}{m_i+m_k}\right)\left[1-\frac{1}{2}\left(1+ \varepsilon_{ki}
\right)\frac{\gamma_{ik}m_k+m_i}{\left(m_i+m_k\right)} \right] 
\end{eqnarray}
do not depend on time.
The inverse characteristic time of granular temperature decay is equal for all species present in the system
\begin{equation}
\tau_0^{-1}=\frac{1}{2}\xi_k(0)=\frac{1}{2}\sum_{i=1}^N\hat{\xi}_{ki}\,.
\end{equation}
Performing the integration of Eq.~(\ref{R2}), we obtain the MSD:
\begin{eqnarray}\nonumber
&&\left< R_k^2(t)\right>=6D_k(0)\tau_0\log\left(1+\frac{t}{\tau_0}\right)+\\
&&+6D_k(0)\tau_{vk}(0)\left(\left(1+\frac{t}{\tau_0}\right)^{-\beta_k}\!-1\right)\label{R2anal}
\end{eqnarray}
with $\beta_k=\tau_0/\tau_{v,k}(0)$.
At long times, $t\gg\tau_0$ the first diffusive term becomes dominant:
\begin{eqnarray}
\left< R_k^2(t)\right>=6D_k(0)\tau_0\log\left(1+\frac{t}{\tau_0}\right)\,.
\label{R2anallong}
\end{eqnarray}
This expression coincides with the one obtained in \cite{garzointruder24}. However, it is obtained in a long time limit, practically unreachable in real experiments.

The initial velocity correlation time may be obtained from Eqs.~(\ref{tausum}-\ref{tauvconst}):
\begin{eqnarray}\nonumber
&&\tau_{v,k}^{-1}(0)= \frac{2\sqrt{2T_k(0)\pi}}{3}\sum_{i=1}^N n_i\sigma_{ki}^{2}g_{2}(\sigma_{ki})\frac{m_i}{\sqrt{m_im_k}}\times\\&&\times \frac{\left(m_i+\gamma_{ik}m_k\right)^{3/2}}{\left(m_i+m_k\right)^2}\left(1+\varepsilon_{ki}\right)^2\,.
\end{eqnarray}

\begin{figure}\centerline{\includegraphics[width=0.95\columnwidth]{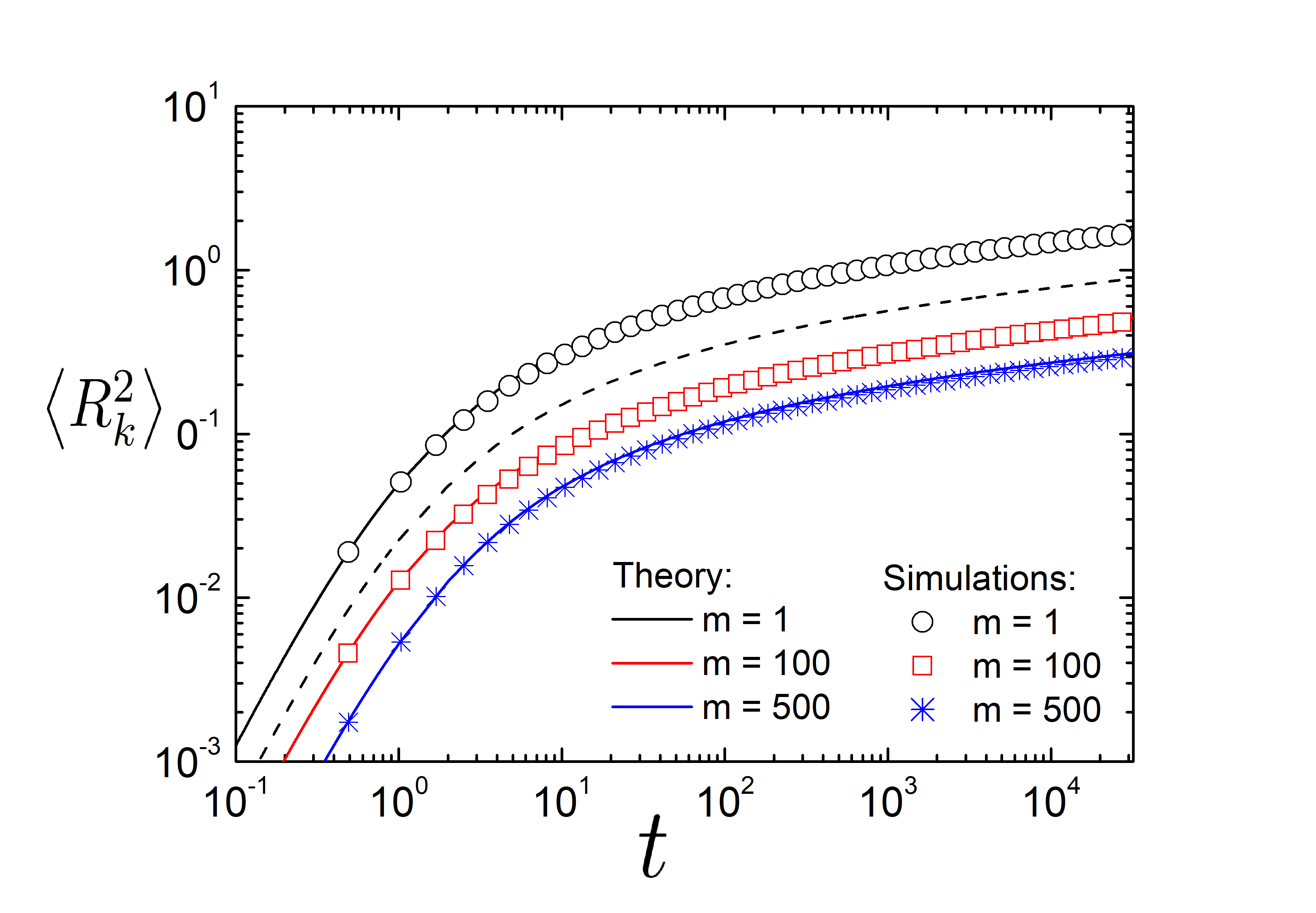}}\caption{Time dependence of partial MSDs in a ternary granular mixture. The restitution coefficient $\varepsilon=0.5$. The partial number densities of particles are $n_1=n_2=n_3=0.1$ and there are $N_1=N_2=N_3=10^4$ particles of each size. The masses of species are $m_1=1$, $m_{2}=100$, $m_3 = 500$ the diameters $\sigma_1=1$, $\sigma_{2}=100^{1/3}$, $\sigma_{3}=500^{1/3}$. The initial granular temperatures are $T_1(0) = 0.048$, $T_2(0) = 1.1$, $T_3(0) = 1.845$, the temperature relaxation time $\tau_0=1.83$. Symbols correspond to the results of DSMC simulations. The black dashed line corresponds to the total MSD $\left< R^2(t)\right>$ (Eq.~(\ref{MSD})).} \label{Gm1m100m500Os}
\end{figure}

\subsection{MSD: results and discussion}

The time evolution of MSD is in a binary mixture is illustrated in Fig.~\ref{Grealnew}. There is big amount of lighter particles of mass $m_1=1$ (gas particles) with number density $n_1=0.1$ and much smaller amount ($n_2=0.001$) of more massive particles (intruders) of mass $m_2=100$. Both types of particles are produced of the same material. Our measurement starts when the ratio of granular temperatures has reached a constant value, and the initial granular temperatures are equal to $T_1(0) = 0.554$, $T_2(0) = 45.58755$. In such a way, the average granular temperature is equal to unity:
\begin{equation}\label{eq:ratio_at_1_crit}
 \left\langle T (0)\right\rangle=\frac{n_1 T_1(0)+n_2 T_2(0)}{n_1+n_2}=1.
\end{equation} 
The interactions between the intruders themselves are rare and they rarely collide with each other. The trajectories of particles of mass $m_2$ are not significantly perturbed by collisions with lighter particles $m_1$; therefore the intruders with a higher mass move faster. Symbols in Fig.~\ref{Grealnew} correspond to the results of DSMC simulations; the detailes of the simulations are provided in the next Section III. The dashed line in Fig.~\ref{Grealnew} shows the MSD given by only the diffusive term in Eq.~(\ref{R2anallong}). One can see that this expression fits the simulations only at relatively long times, $t\gg\tau_0$. For larger particles, the difference between the full (Eq.~\ref{R2anal}), and simplified (Eq.~\ref{R2anallong}) expressions becomes more significant.

The ternary granular mixture, in which equal amounts of different species are present in the system, is illustrated in Fig.~\ref{Gm1m100m500Os}. Now the particles of larger mass also intensively interact with each other, and increase of mass of particles leads to slowing down of their motion. The dashed line corresponds to total MSD $\left< R^2(t)\right>$ (Eq.~(\ref{MSD})), the solid lines are provided by Eq.~(\ref{R2anal}). The nice agreement with DSMC results shows that this analytical expression is valid also for multicomponent granular mixtures.

\begin{figure}\centerline{\includegraphics[width=0.95\columnwidth]{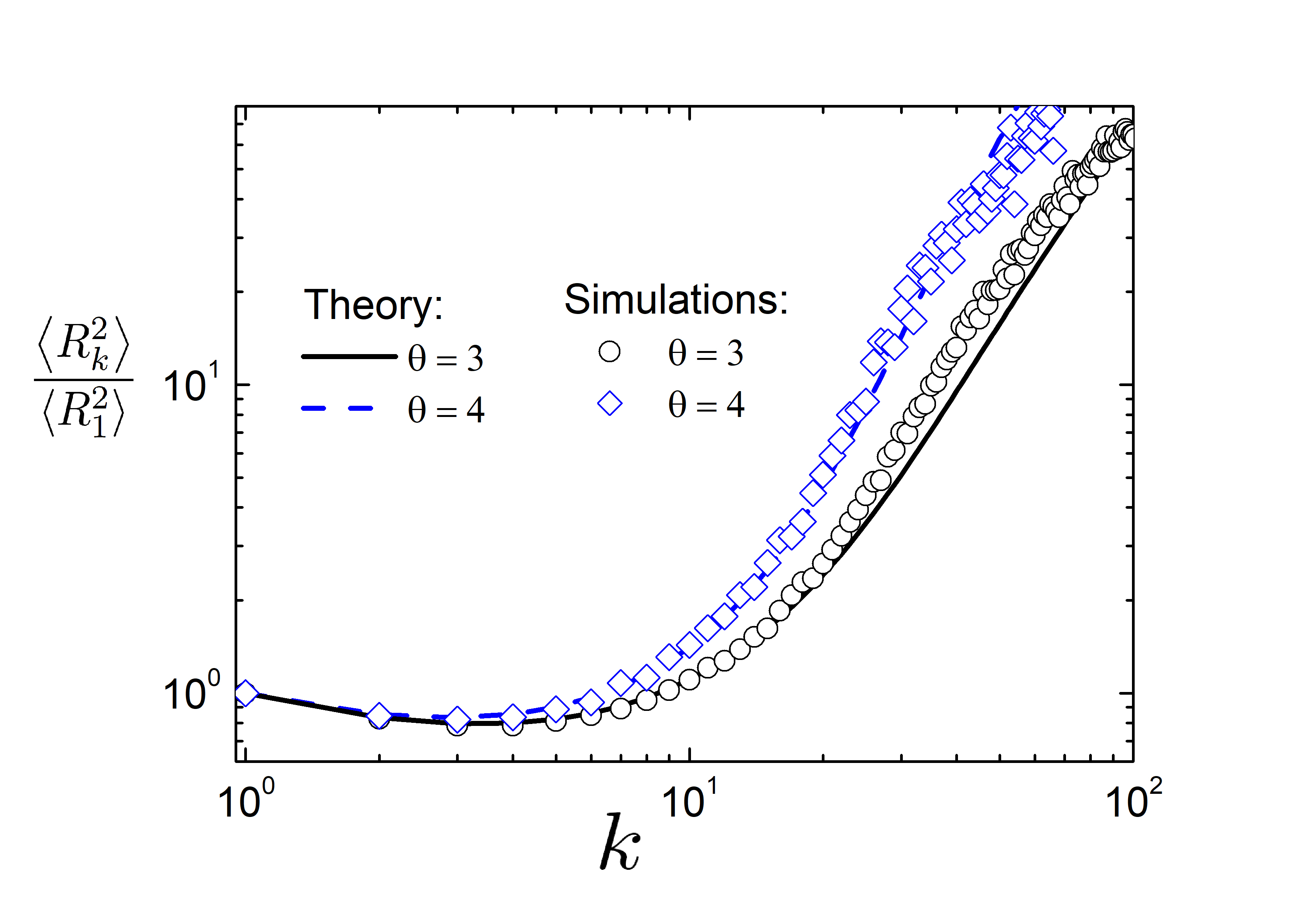}}\caption{Fraction of partial MSDs $\left<R_k^2\right>/\left<R_1^2\right>$ for different values of $k = m_k/m_1$ in a mixture of granular particles with number density $n_k = n_1k^{-\theta}$, $\theta=3,4$, $n_1 = 0.1$ number of monomers $N_1 = 10^7$, time $t = 10^9$ and constant restitution coefficient $\varepsilon = 0.5$. Lines correspond to the result of numerical integration of Eq.~(\ref{R2}), symbols denote the results of MC simulations} \label{GRk}\end{figure}

Now let us consider a highly polydisperse granular system with a discrete distribution of particle masses. Let the smallest particle mass be $m_1 = 1$ and the masses of the other particles be $m_k = km_1$, where $k = 1, 2, …N$ are integers and $N$ is the total number of different species in the system. The mixtures contain a large amount of species $N\gg 1$. 
We assume that the number densities are distributed according to $n_k=n_1k^{-\theta}$. In this case, the granular temperature distribution scales according to $T_k\sim k^{5/3}$ \cite{lev} and the MSD has the following size-dependence at $k\gg 1$: $\left< R^2_k\right>\sim k^{5/3}$ for $k\gg 1$, $t\gg\tau_0$ \cite{anna2024}. The ratio $\left< R^2_k\right>/\left< R^2_1\right>$ is plotted in Fig.~\ref{GRk}. One can see that, in the beginning, the ratio first decreases, then reaches its minimum value, and then starts to increase again. It is due to the fact that the ratio of granular temperatures grows very slowly in the beginning, according to $T_k\sim k^{\alpha}$ with $\alpha<1$ \cite{lev}, and the characteristic velocity of particles of size $k$ decreases with increasing of $k$. Then, $\alpha$ takes values larger than unity, and the characteristic velocity increases again. An efficient algorithm, described in Section III, allows considering diffusion in very large and highly polydisperse systems. Good agreement between the simulations and theory is observed, although large particles require a large amount of time to reach the theoretically predicted MSD ratios.

\subsection{Time-averaged mean-squared displacement}

The time-averaged MSD is introduced to evaluate the time series in experiments and simulations \cite{jeon,sok, eli,m1,golding,wang}:
\begin{equation}
\left\langle\overline{\delta_k^2(\Delta)}\right\rangle=\frac{1}{t-\Delta}\int_0^{
t-\Delta}\left\langle\left[\mathbf{R}_k(t^{\prime}+\Delta)-\mathbf{R}_k(t^{\prime})
\right]^2\right\rangle dt'\,.
\label{taMSD}
\end{equation}  
Here $\Delta$ is the lag time, and the angular brackets denote the average
over all traces of particles of type $k$. For an ergodic system, such as an ideal gas with a unit
restitution coefficient corresponding to normal particle diffusion, the ensemble and
time-averaged
MSDs are equivalent at any time, $\langle R_k^2(\Delta)\rangle=\langle\overline{\delta_k^2(\Delta)}
\rangle$ \cite{jeon,sok, eli,m1}. In contrast, several systems characterized by anomalous
diffusion with power-law MSD $\langle R_k^2(t)\rangle\simeq t^{\alpha}$ ($\alpha\neq
1$) or a corresponding logarithmic growth of the MSD, are non-ergodic and display the
disparity $\langle R_k^2(\Delta)\rangle\neq\langle\overline{\delta_k^2(\Delta)}
\rangle$ \cite{jeon,sok, eli,m1,sinai,metzlerprl08,schulz,glass}.

The partial time-averaged MSD for species $k$ in a multicomponent granular gas may be obtained performing the steps analogous to the monodisperse granular system \cite{annapccp}:
\begin{eqnarray}
\label{delta01}
\left<\overline{\delta_k^2(\Delta)}\right>=\left<\overline{\delta_0^2(\Delta)}\right>
+\Xi(\Delta).
\end{eqnarray}
Here the first term is equal to
\begin{eqnarray}
\nonumber
&&\left\langle\overline{\delta_0^2(\Delta)}\right\rangle=\frac{6D_k(0)\tau_0}{t-\Delta}\Big[(t+\tau_0)\log(t+\tau_0)-\\
\nonumber
&&-(\Delta+\tau_0)\log(\Delta+\tau_0)-\\
\nonumber
&&-(t-\Delta+\tau_0)\log(t-\Delta+\tau_0)+\tau_0\log\tau_0\Big].
\label{delta00}
\end{eqnarray}
For $\tau_0\ll\Delta\ll t$ it tends to
\begin{equation}
\left\langle\overline{\delta_0^2(\Delta)}\right\rangle\sim\frac{6D_k(0)\tau_0\Delta}
{t}\left[\log\left(\frac{t}{\Delta}\right)+1\right].
\end{equation}
The second term in Eq. (\ref{delta01}) may be obtained performing the numerical integration
\begin{eqnarray}\nonumber
\Xi(\Delta)=\frac{6D_k(0)\tau_{v,k}(0)}{t-\Delta}\int_{0}^{t-\Delta}dt^{\prime}\left[
\left(1+\frac{\Delta}{t^{\prime}+\tau_0}\right)^{-\beta_k}-1\right]\,.
\end{eqnarray}

\subsection{Binary granular system: tracer limit}

Let us consider a binary granular system of particles with masses $m_1$ (gas particles) and $m_2$ (intruder particles). Let $\mu=m_2/m_1$ be the ratio of the mass of the particles. We assume that the number density $n_2\ll n_1$. In this case, the presence of the intruder does not affect the evolution of the surrounding granular gas, and the interaction between the intruders may be neglected. By formally setting $n_2=0$ we obtain a single intruder surrounded by gas particles. 

The evolution of granular gas particles occurs according to Haff's law  (Eq.~\ref{TkHaff}) \cite{haff} for $k=1$ with relaxation time \cite{book}
\begin{equation}
\tau_0 = \frac{6}{1-\varepsilon^2}\tau_c(0)\,.
\end{equation}
The velocity correlation time is equal to
\begin{equation}
\tau_{v,1}(t)=\tau_c(t)\frac{6}{\left(1+\varepsilon\right)^2}\,.
\end{equation}
The MSD may be obtained from the Eq.~(\ref{R2anal}) with $k=1$ and is the same as in the case of a monodisperse granular gas \cite{annapccp}:
\begin{eqnarray}\nonumber&&\left< R_1^2(t)\right>=36D_1(0)\tau_c(0)\left(\frac{1}{1-\varepsilon^2}\log\left(1+\frac{t}{\tau_0}\right)\right.\\&&+\left.\frac{1}{\left(1+\varepsilon\right)^2}\left(\left(1+\frac{t}{\tau_0}\right)^{-\left(1+\varepsilon\right)^2/\left(1-\varepsilon^2\right)}\!-1\right)\right)\,.
\label{R2gas}\end{eqnarray}

\begin{figure}\centerline{\includegraphics[width=0.95\columnwidth]{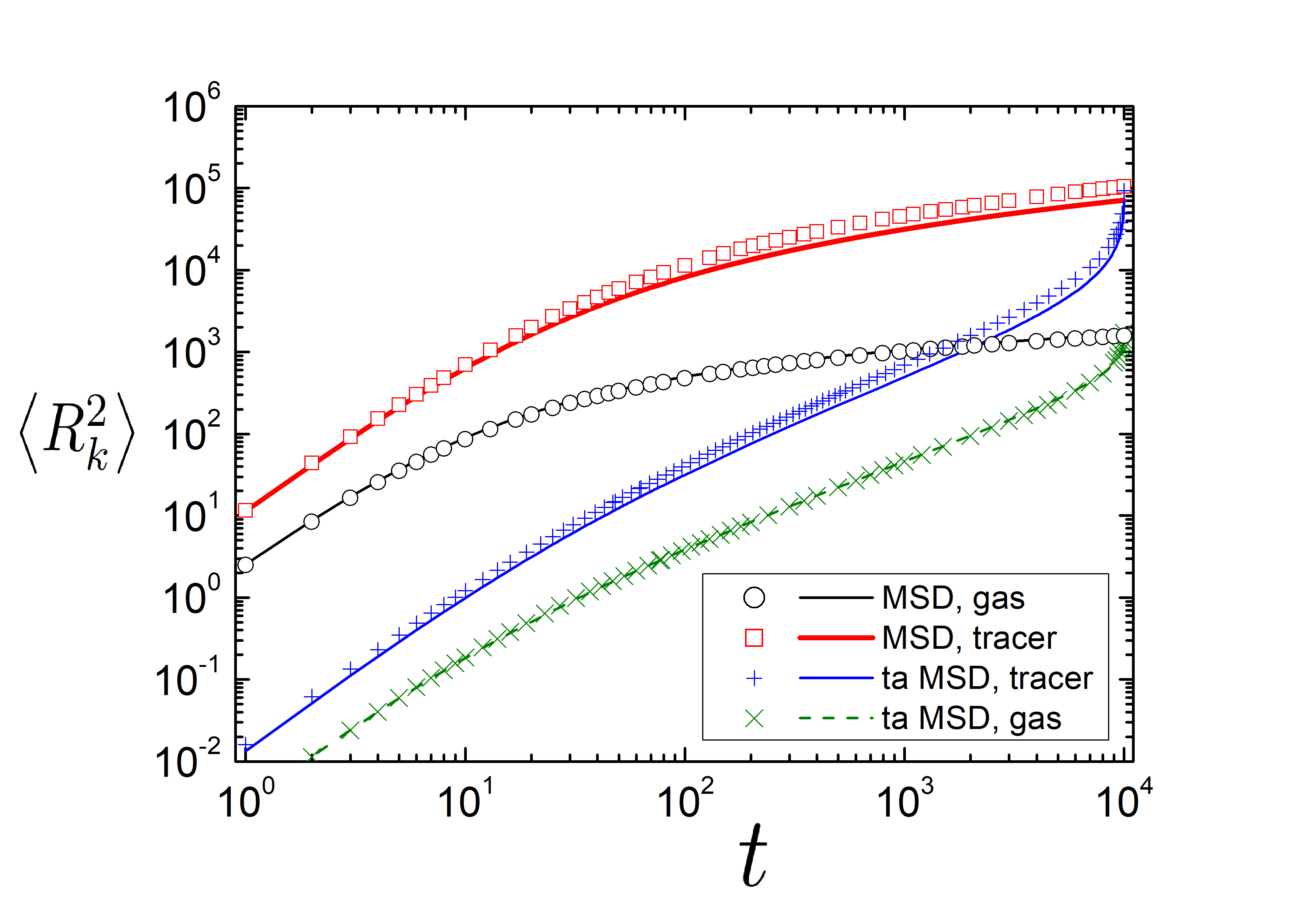}}\caption{MSD and time-averaged MSD in a binary granular mixture in a tracer limit: $n_1=0.1, n_2=0$. The initial granular temperatures are $T_1(0)=1$, $T_2(0) = 412.93$. The masses of species are $m_1=1$, $m_{2}=100$, the diameters $\sigma_1=1$, $\sigma_{2}=100^{1/3}$. The granular temperature relaxation time $\tau_0=11.3$. The velocity correlation time of tracers $\tau_{v,2}=1212$. Restitution coefficient $\varepsilon=0.5$. Symbols denote the results of MC simulations.}
\label{Gtracer}\end{figure}

The MSD of the intruder particles takes the form of Eq.~(\ref{R2anal}) with the initial velocity correlation time
\begin{eqnarray}
&&\tau_{v,2}(0)=\tau_c(0)\hat{\tau}_{v,21}\,,\\
&&\hat{\tau}_{v,21}=\frac{6\sigma_1^2\left(1+\mu\right)^2\gamma_{21}\sqrt{\mu}}{\sqrt{2}\sigma_{12}^2(\mu+\gamma_{21})^{3/2}(1+\varepsilon_{12})^2}\,.
\end{eqnarray}
For $\mu\gg 1$, $\varepsilon_{22}\to 1$ the temperature ratio $\gamma_{21}$ may be found explicitly \cite{book, garzoreview, brey}:
\begin{eqnarray}
\gamma_{21} &=&\frac{T_2}{T_1}=\frac{1+\varepsilon_{12}}{2\left(1-b\right)}\,\\
b &=& \frac{\mu}{2\sqrt{2}}\frac{1-\varepsilon_{11}^2}{1+\varepsilon_{12}}\frac{\sigma_{12}^2}{\sigma_2^2}\frac{g_2(\sigma_{12})}{g_2(\sigma_2)}\,.
\end{eqnarray}
Both the MSDs and time-averaged MSD for intruder particles and surrounding gas particles are shown in Fig.~\ref{Gtracer}. In the tracer limit, the magnitude of the MSD is significantly higher, as for granular mixtures even at low but finite number densities (Fig.~\ref{Gtracer}). The agreement of both quantities with DSMC simulations are good.  


\section{Monte Carlo simulations}

\subsection{Collision rates}

To simulate these systems, we use the standard DSMC approach \cite{compbook, Bird}, which was modified to perform simulations of polydisperse mixtures faster.

The main idea of our approach is to split the selection of particles into two steps. First, their sizes are selected (according to the total collision rates $C_{ik}$ of particles of types $i$ and $k$), and then particles $j$ and $l$ of types $i$ and $k$ are selected randomly with a uniform probability. Finally, rejection sampling is performed, so that the final collision rates $C_{ik}^{jl}$ are correct. We assume that the collisions are pairwise and neglect the possible simultaneous collisions of multiple particles, which do not occur in rarefied granular systems.

The collision rates $C_{ik}^{jl}$ between particles with numbers $j$ and $l$ of masses $m_i = i m_1$ and $m_k = k m_1$ and velocities $\textbf{v}_i^j$ and $\textbf{v}_k^l$ can be determined from the so-called collision cylinder \cite{book}:
\begin{equation}
  C_{ik}^{jl} = \pi \sigma_{ik}^2 \left| \left( \textbf{v}_i^j - \textbf{v}_k^l, \textbf{e} \right) \right| / V\,.
\end{equation}
Here $\sigma_{ki}=\left(\sigma_k+\sigma_i\right)/2$. 

We set the upper limit $C_{ik}$ for the rates $C_{ik}^{jl}$, where $C_{ik}$ does not directly depend on $j$ and $l$:
\begin{equation}
  C_{ik}^{jl} \leq C_{ik} = \pi \sigma_{ik}^2 \left( \max\limits_j \left| \textbf{v}_i^j \right| + \max\limits_l \left| \textbf{v}_k^l \right| \right) / V\,.
\end{equation}
The maximal velocities are saved in advance and updated whenever the maximum increases (or during output when we go through all arrays of particles).

The DSMC algorithm can be written in five steps:
\begin{enumerate}
    \item Advance time using total collision rate: 
    \begin{equation}\label{eq:LowRankExpTrial}
        t := t - \log \left( {\rm rand} \left( 0, 1 \right] \right) / \sum\limits_{i,k} C_{ik} N_i N_k.
    \end{equation}
    \item Select sizes $i$ and $k$ according to the probabilities
    \begin{equation}\label{eq:Pik}
      P_{ik} = \frac{C_{ik} N_i N_k}{\sum\limits_{p,q} C_{pq} N_p N_q}.
    \end{equation}
    \item Select particles $j$ (from $1$ to $N_i$) and $l$ (from $1$ to $N_k$) of sizes $i$ and $k$ uniformly at random.
    \item Generate collision direction $\textbf{e}$ and accept collision with probability \begin{equation}\label{eq:Pikjl}
    P_{ik}^{jl} = \frac{\left| \left( \textbf{v}_i^j - \textbf{v}_k^l \right) \cdot \textbf{e} \right|}{\max\limits_j \left| \textbf{v}_i^j \right| + \max\limits_l \left| \textbf{v}_k^l \right|}.
    \end{equation}
    Otherwise, go to step 1.
    \item Update the velocities according to 
       \begin{eqnarray}
  \label{v1v2} 
  &&{\bf v}_{k}^{\,\prime} = {\bf v}_{k} - \frac{m_i}{m_i+m_k}\left(1+\varepsilon_{ki}\right)({\bf v}_{ki} \cdot {\bf e})\,{\bf e} \,,\\
  &&{\bf v}_{i}^{\,\prime} = {\bf v}_{i} + \frac{m_k}{m_i+m_k}\left(1+\varepsilon_{ki}\right)({\bf v}_{ki} \cdot {\bf e})\,{\bf e} \,.
  \nonumber
 \end{eqnarray}
   
\end{enumerate}

The acceptance rate $P_{ik}^{jl}$ ensures that the equality $C_{ij} P_{ik}^{jl} = C_{ik}^{jl}$ holds so that the final collision rates are exactly the same as they need to be.

As mentioned in \cite{Bird}, simply choosing particles uniformly at random when the size ratio is high leads to a significant performance degradation. If the sizes are quickly selected in advance, there is no such problem because they do not appear in the acceptance rates $P_{ij}^{kl}$, Eq.~(\ref{eq:Pikjl}). One can even create arrays of size 1 for each particle $j$ (as if all particles had different sizes) and have collision rates kernel $C_{jk}$ of size $N$ by $N$, which allows getting rid of step (iii) and using the exact velocity of particle $j$ in Eq.~(\ref{eq:Pikjl}).

To quickly select the sizes, instead of calculating the probabilities $P_{ik}$ from Eq.~(\ref{eq:Pik}) directly, we use the low-rank method analogous to \cite{mediag}. First, we use the symmetry of $C_{ik} N_i N_k$,
\begin{equation}
C_{ik} N_i N_k = A_{ik} + A_{ki}, \quad A_{ik} = \pi \sigma_{ik}^2 \max\limits_j \left| \textbf{v}_i^j \right| / V
\end{equation}
and then observe that $A_{ik}$ is a rank 3 matrix, since $\sigma_{ik} = \left(\sigma_i + \sigma_k\right)/2$:
\begin{eqnarray}
  A_{ik} & = \frac{\pi \max\limits_j \left| \textbf{v}_i^j \right|}{4V} \left( \sigma_i^2 \cdot 1 +  2\sigma_i \cdot \sigma_k + 1 \cdot \sigma_k^2 \right) \\
  & = u_i^{(1)} v_k^{(1)} + u_i^{(2)} v_k^{(2)} + u_i^{(3)} v_k^{(3)}.
  \end{eqnarray}
Each term here allows for the separation of variables (i.e., the separation of indices $i$ and $k$) and allows us to select sizes, according to the vectors $u_i^{(r)}$ and $v_k^{(r)}$, $r \in \left\{ 1, 2, 3 \right\}$. One of the three terms can be selected using the probabilities $P_r = \frac{\sum\limits_i u_i^{(r)} \sum\limits_k v_k^{(r)}}{\sum\limits_r \sum\limits_i u_i^{(r)} \sum\limits_k v_k^{(r)}}$. Once $r$ is selected, we use the probabilities $P_i^{(r)} = \frac{u_p^{(r)}}{\sum\limits_p u_p^{(r)}}$ and $P_k^{(r)} = \frac{v_k^{(r)}}{\sum\limits_q v_q^{(r)}}$. To use them quickly in practice, we construct segment trees on $u^{(r)}$ and $v^{(r)}$: this data structure allows performing searches and updates (including the update of the total sums used in $P_i^{(r)}$, $P_k^{(r)}$ and $P_r$) in $O \left( \log M \right)$ operations (where $M$ is the number of different cluster sizes), leading to the total logarithmic cost of the whole algorithm. Indeed, velocity distributions have exponential tails; thus, $k$ and $l$ selection costs $O \left( \log N \right)$ on average, which is also logarithmic. Naturally, with only two or three different sizes in the system, the segment trees are not required.

Unlike the methods from \cite{mediag} and \cite{mesmollowrank}, which were developed for (generalized) Smoluchowski equations, we have an additional rejection step \eqref{eq:Pikjl}. This allows us to eliminate the Maxwell distribution assumption (which is not exactly satisfied for cooling granular gases; the velocity distribution tail is known to be exponential \cite{book}) by keeping track of the velocities of each individual particle, which will also be needed to calculate the displacements. This approach not only requires no assumptions about velocity distributions (initial distribution at $T = 1$ can be obtained by first starting at some point, where the temperature was higher), but also allows skipping the derivation of temperature-dependent equations \cite{nature}, which can be very cumbersome, especially if one uses realistic restitution coefficients. The simulations of systems with velocity-dependent restitution coefficients are described in Appendix A.

To calculate the mean displacement, we keep track of the displacement $\textbf{R}_k^l$ for each individual particle $l$ of each individual size $k$ and then average $\textbf{R}_k^l$ over $l$ during the output.

Each time a particle $l$ of size $k$ participates in a collision, we update its relative displacement as
\begin{equation}
  \textbf{R}_k^l := \textbf{R}_k^l + \textbf{v}_k^l \left( t - t_k^l \right),
\end{equation}
where $\textbf{v}_k^l$ is the pre-collision velocity and $t_k^l$ is the system time of the last collision (which we keep track of for each particle). When we save the current displacements in a file at time $t$, we also add $\textbf{v}_k^l \left( t - t_k^l \right)$ to the saved displacements without updating $\textbf{R}_k^l$ or $t_k^l$.

To derive the partial MSD $\left< R_k^2(t)\right>$, the average over displacements $\textbf{R}^2_k$ of all particles of species $k$ at time $t$ is calculated. A comparison of our DSMC simulations with the analytical results is shown in Figs. 1-5, and a good agreement can be observed.

\begin{figure}\centerline{\includegraphics[width=0.95\columnwidth]{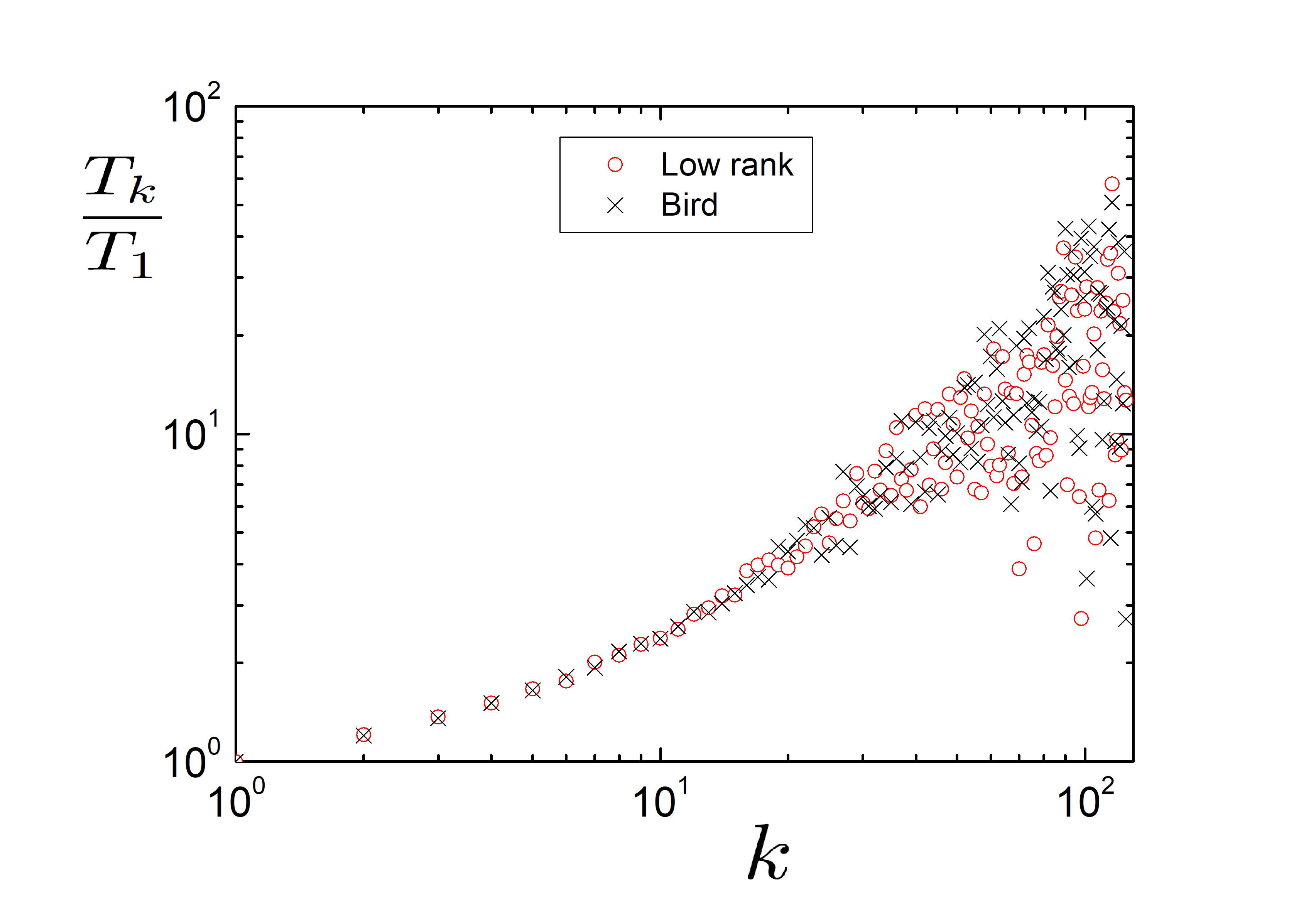}}
\caption{Temperatures $T_k$ of species for different values of $k = m_k/m_1$ in a mixture of granular particles with number densities $n_k = n_1k^{-3}$, $n_1 = 0.1$ number of monomers $N_1 = 10^6$, time $t = 100$ and constant restitution coefficient $\varepsilon = 0.5$. Symbols denote the results of different DSMC simulations both in terms of low rank and Bird's method.} \label{fig:compare}\end{figure}

\subsection{Comparison with Bird's method}

Here, we compare our technique with Bird's method. When generalized to the case of different particle sizes, the following collision rate exists between particles $j$ and $l$ with diameters $\sigma^{(j)}$ and $\sigma^{(l)}$, which is exactly defined using the collision cylinder:
\begin{equation}
  C^{jl} = \pi \left( \frac{\sigma^{(j)} + \sigma^{(l)}}{2} \right)^2 \left| \left( \textbf{v}_i^j - \textbf{v}_k^l, \textbf{e} \right) \right| / V.
\end{equation}
Velocities can be bound by some multiple of system temperature \cite{book} or (which is more accurate) by the current maximum velocity in the system. Particles are then selected at random and accepted with a probability that is proportional to the ratio of the real collision rate (with random collision direction $\vec e$) and maximum possible collision rate. When the sizes are equal, we obtain an acceptance probability similar to \eqref{eq:Pikjl}; however, in general, we have the ratio
\begin{equation}\label{eq:maxrat}
  P_{\rm exact}^{jl} = \frac{\left( \frac{\sigma^{(j)} + \sigma^{(l)}}{2} \right)^2 \left| \left( \textbf{v}^j - \textbf{v}^l \right) \cdot \textbf{e} \right|}{\max\limits_{j,l} \left( \frac{\sigma^{(j)} + \sigma^{(l)}}{2} \right)^2 \left( \left| \textbf{v}^j \right| + \left| \textbf{v}^l \right| \right)}.
\end{equation}
Because calculating the maximum here requires checking all $M^2$ pairs of sizes, we can simplify this by opening the last brackets and assuming that the second particle in the pair has the maximum size $M$:
\begin{equation}\label{eq:PjlBird}
\begin{aligned}
  P^{jl} & = \frac{\left( \frac{\sigma^{(j)} + \sigma^{(l)}}{2} \right)^2 \left| \left( \textbf{v}^j - \textbf{v}^l \right) \cdot \textbf{e} \right|}{\max\limits_{j} \left( \frac{\sigma^{(j)} + \sigma_M}{2} \right)^2 \left| \textbf{v}^j \right| + \max\limits_{l} \left( \frac{\sigma_M + \sigma^{(l)}}{2} \right)^2 \left| \textbf{v}^l \right|} \\
  & = \frac{\left( \frac{\sigma^{(j)} + \sigma^{(l)}}{2} \right)^2 \left| \left( \textbf{v}^j - \textbf{v}^l \right) \cdot \textbf{e} \right|}{2\max\limits_{j} \left( \frac{\sigma^{(j)} + \sigma_M}{2} \right)^2 \left| \textbf{v}^j \right|},
\end{aligned}
\end{equation}
where it is easily verified that the denominator cannot be more than 2 times higher than the exact maximum in \eqref{eq:maxrat}. The possibility of these rare collisions with large particles makes Bird's method slow when $M$ is large, as this leads to a low acceptance rate. The bound
\begin{equation}
  C_{\max} = 2\max\limits_{j} \left( \frac{\sigma^{(j)} + \sigma_M}{2} \right)^2 \left| \textbf{v}^j \right|
\end{equation}
can be quickly updated when the new velocity of a particle exceeds the maximum (or once every $N$ collisions to account for the decrease in the average kinetic energy). Because $C_{ik} \leqslant C_{\max}$ for all $i$ and $k$, the low-rank method always has fewer rejections, which makes it faster. If most of the collisions occur between particles of size $i \sim j \sim 1$ and they have the maximum velocity, one can expect about $C_{\max} / C_{11} \sim \sigma_M^2/\sigma_1^2 \sim M^{2/3}$ times more rejections between collision events.

As mentioned previously, the low-rank method produces the same (correct) collision rates. Moreover, the time steps are also the same. In Bird's method, the time step between trials is
\begin{equation}
  \Delta t_{\rm trial} = - \log \left( {\rm rand} \left( 0, 1 \right] \right) / \left( N^2 C_{\max} \right),
\end{equation}
assuming ordered pairs are selected, and one then rejects the case  when the same particle is selected twice. This produces an exponential distribution with the expectation $1 / \left( N^2 C_{\max} \right)$. After accounting for the total acceptance probability by averaging over \eqref{eq:PjlBird}, it turns into exponential distribution with expectation
\begin{equation}\label{eq:BirdExp}
\begin{aligned}
\mathbb{E} \Delta t_{\rm coll} & = \frac{1}{N^2 C_{\max} \sum\limits_{j,l} \frac{P_{jl}}{N^2}} \\
& = \frac{1}{C_{\max}} \cdot \frac{2\max\limits_{j} \left( \frac{\sigma^{(j)} + \sigma_M}{2} \right)^2 \left| \textbf{v}^j \right|}{\sum\limits_{j,l} \left( \frac{\sigma^{(j)} + \sigma^{(l)}}{2} \right)^2 \left| \left( \textbf{v}^j - \textbf{v}^l \right) \cdot \textbf{e} \right|} \\
& = \frac{1}{\sum\limits_{j,l} C^{jl}}.
\end{aligned}
\end{equation}
Similarly, in the low-rank method, we have an exponential distribution with expectation as in \eqref{eq:LowRankExpTrial} for each trial and with the expectation
\begin{equation}
 \mathbb{E} \Delta t_{\rm coll} = \frac{1}{\sum\limits_{i,k} C_{ik} N_i N_k \sum\limits_{j,l} \frac{P_{ik}^{jl}}{N_iN_k}} = \frac{1}{\sum\limits_{i,k,j,l} C_{ik}^{jl}} = \frac{1}{\sum\limits_{j,l} C^{jl}}
\end{equation}
after each accepted collision, which coincides with \eqref{eq:BirdExp}.

Thus, unlike, for example, Nanbu's method, where several collisions are performed simultaneously, leading to accuracy loss because there are no updates of collision rates between the selected collisions \cite{BirdNanbuCompare}, our method produces exactly the same collision rates as Bird's method.

The equivalence of the suggested approach to Bird's method is shown in table \ref{tab:compare} and Fig. \ref{fig:compare}. We ran both methods for the number density distribution $n_k = n_1 k^{-3}$, $n_1 = 0.1$ and $N_1 = 10^6$ monomers up to $t = 100$ with $T(0) = 1$ and the initial Maxwell distribution. We see that the difference between the two methods is of the order of the stochastic noise in both of them $1/\sqrt{N} \sim 0.001$. On the other hand, calculations using the low-rank structure allow the selection of pairs much faster, without losing any information in the process. The difference becomes even larger for larger computation times when the temperatures of large clusters increase further, which leads to a corresponding increase in the maximum collision rate and the number of rejections. With Bird's method and the parameters of Fig. \ref{GRk}, it would take days or weeks to reach times of the order $t \sim 10^9$ required to see the convergence of partial MSDs.

\begin{table}[ht]
\caption{Comparison between low-rank and classical DSMC methods. Simulations were performed for $N_1 = 10^6$ monomers with density $n_1 = 0.1$ and density of other species $n_k = n_1 k^{-3}$, where the numbers of particles are rounded to the nearest integer. Initial temperature $T(0) = 1$ and initial velocity distribution is Maxwell distribution. Restitution coefficient $\varepsilon = 0.5$.}\label{tab:compare}
\begin{center}
\begin{tabular}{|l|c|c|c|}
\hline
\multirow{2}{*}{Method} & Collisions & Total kinetic & Computation \\
& performed & energy reached & time, sec \\
\hline
Low-rank DSMC & $12072722$ & $1.233 \cdot 10^5$ & $28$ \\
\hline
Bird DSMC & $12067760$ & $1.235 \cdot 10^5$ & $1128$ \\
\hline
\end{tabular}
\end{center}
\end{table}
As previously mentioned, Bird's DSMC is currently the preferred and most widely used DSMC method, because of its lower complexity and better accuracy compared, for example, with Nanbu's method. 
Event-driven molecular dynamics is also widely used for the investigation of granular systems \cite{compbook}. It is very different in nature and is known to be much slower than DSMC. When the molecular chaos hypothesis is satisfied \cite{book}, the DSMC methods are always preferred.

Comparison of low-rank and Bird DSMC methods and code for various methods for Smoluchowski equations are available at https://github.com/RodniO/Low-rank-Monte-Carlo

\section{Conclusions}

We developed an efficient numerical algorithm for the investigation of diffusion coefficients and mean squared displacements in highly polydisperse granular systems and obtained very good agreement between our simulations and analytical results. Variations in the size, mass, restitution coefficient and number density of particles may significantly affect their motility. We also discussed ergodicity breaking in polydisperse granular systems by deriving and comparing ensemble and time-averaged MSDs in dilute binary granular mixtures. We have obtained an explicit analytic expression for the MSD that is valid both at short and long times in systems with an arbitrary number of particles of different sizes and masses. 
Our results may be helpful in industrial applications involving different types of granular materials, to understand the motion of the constituents of interstellar dust clouds, planetary rings, and other astrophysical objects.

\section{Acknowledgement} A.S.B. thanks Ralf Metzler for the fruitful discussions.

\begin{figure}\centerline{\includegraphics[width=0.95\columnwidth]{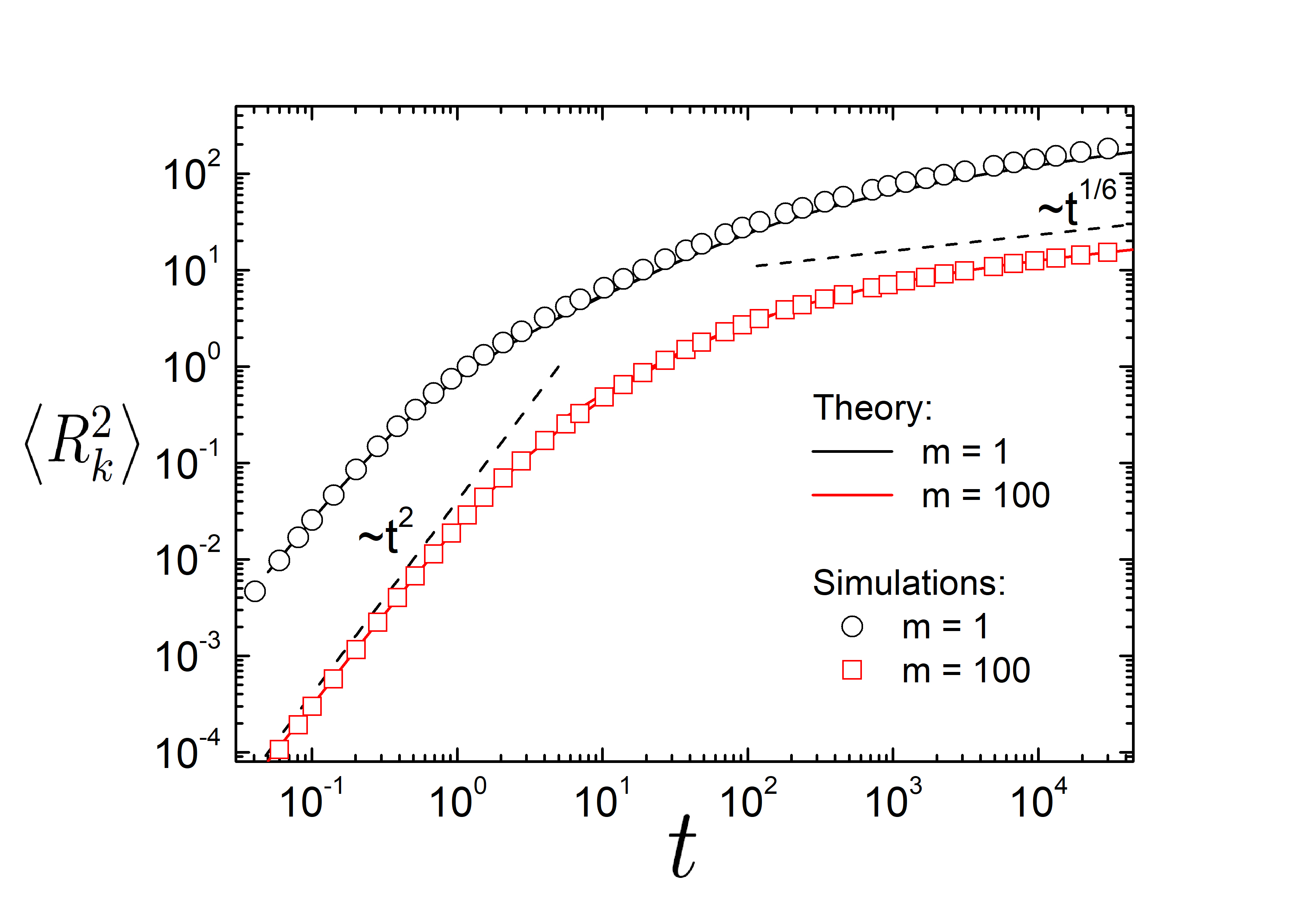}}\caption{MSD in a binary granular mixture with time-dependent restitution coefficient, $A\kappa^{2/5}=0.09$, $m_1=1$, $m_2=100$, $n_1=n_2=0.1$, $\sigma_1=1$, $\sigma_{2}=100^{1/3}$. Symbols correspond to the results of DSMC simulations.} \label{GR2visco}\end{figure}

\appendix

\section{Velocity-dependent restitution coefficient}

The velocity-dependence of the restitution coefficient has the following form \cite{rpbs99,delayed}:
\begin{eqnarray}\label{epsx} 
&&\varepsilon_{ki} = 1 + \sum^{20}_{j=1} h_j \left( A \kappa_{ki}^{2/5} \right)^{j/2} \left |\left
({\bf{v}}_{ki}\cdot\textbf{e}\right )\right |^{j/10}
\end{eqnarray}
Here, $h_k$ are numerical coefficients, and the parameter $A$ characterizes the elastic and dissipative properties of the particle material \cite{bshp96,goldobin}:
\begin{equation}
\label{A} A =
\frac{1}{Y}\frac{\left(1+\nu\right)}{\left(1-\nu\right)}\left(\frac43\eta_1\left(1-\nu+\nu^2\right)+\eta_2\left(1-2\nu\right)^2\right)\,,
\end{equation}
where ${\eta}_1$ and ${\eta}_2$ are viscosity coefficients. Parameter $\kappa_{ij}$ is a function of Young's modulus $Y$, Poisson's ratio $\nu$, and particle size and mass \cite{book,bshp96,rpbs99}.
\begin{eqnarray}
\kappa_{ki}=\frac{1}{\sqrt{2}} \left (\frac{3}{2}\right )^{3/2}\frac{Y}{1-{\nu}^2}\frac{\sqrt{\sigma_{\rm eff}}}{m_{\rm eff}}\,.
\end{eqnarray}
The effective diameter of colliding particles with diameters $\sigma_i$ and $\sigma_j$ is
\begin{equation}
\sigma_{\rm eff}=\frac{\sigma_i\sigma_k}{\sigma_i+\sigma_k}\,.
\end{equation}
The effective mass is equal to 
\begin{equation}
m_{\rm eff}=\frac{m_i m_k}{m_i+m_k}\,.
\end{equation}
The viscoelastic model agreed well with the experimental data for collisions with low impact velocities \cite{rpbs99}.

\begin{figure}\centerline{\includegraphics[width=0.95\columnwidth]{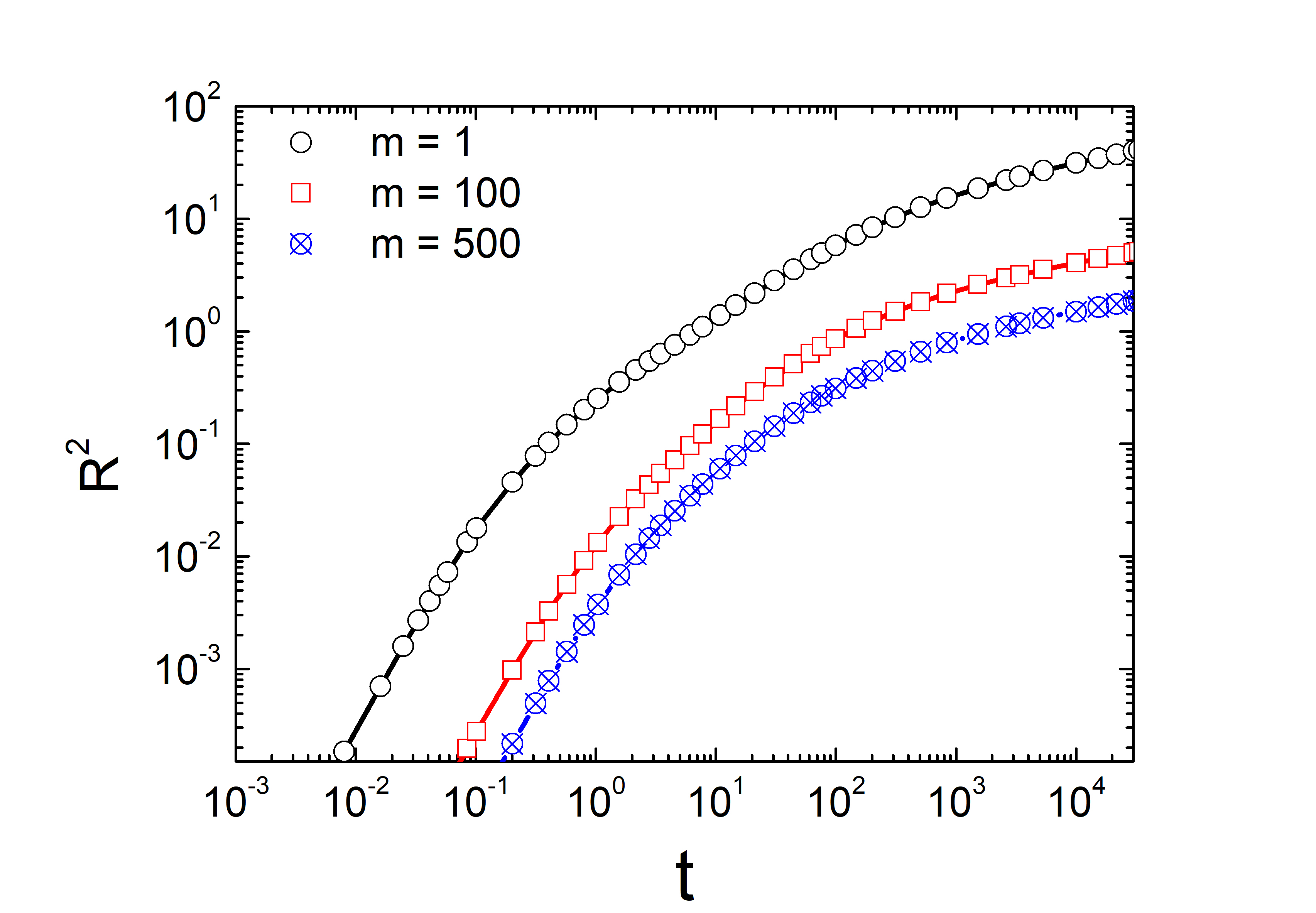}}\caption{MSD in a ternary granular mixture with time-dependent restitution coefficient, $A\kappa^{2/5}=0.09$, $m_1=1$, $m_2=100$, $m_3=500$, $n_1=n_2=n_3=0.1$, $\sigma_1=1$, $\sigma_{2}=100^{1/3}$, $\sigma_{3}=500^{1/3}$. Symbols correspond to the results of DSMC simulations.} \label{GR2visco3}\end{figure}

The MSD is obtained performing the numerical integration of Eq.~(\ref{R2}), where the velocity correlation time is given by Eq.~(\ref{tausum}). The partial inverse velocity correlation times have the form \cite{anna2024}
\begin{eqnarray}\nonumber
&&\tau_{v,ki}^{-1}(t)=\frac{8\sqrt{2\pi}}{3}n_i\sigma_{ki}^{2}g_{2}(\sigma_{ik})m_i\frac{T_km_i+T_im_k}{T_k\left(m_i+m_k\right)^2}\\
&&\times
\left(\frac{T_km_i+T_im_k}{m_im_k}\right)^{1/2}\left(1+\frac{1}{2}\sum_{i}A_iB_i\right)\label{tauvv}
\end{eqnarray}
The evolution of granular temperatures occurs according to Eq.~(\ref{sys}) with cooling rates given by Eq.~(\ref{eq:sys_xyk}), where  \cite{Os}
\begin{eqnarray}\nonumber
&&\xi_{ki}(t) =
\frac{16}{3}\sqrt{2\pi}n_i\sigma_{ki}^{2}g_{2}(\sigma_{ik})
\sqrt{\frac{T_km_i+T_im_k}{m_im_k}}\!\frac{m_i}{m_i+m_k}\times\\&&\nonumber
\left(1-\frac{T_km_i+T_im_k}{T_k\left(m_i+m_k\right)}+\sum_{n}B_n\left(h_n-\frac12\frac{T_km_i+T_im_k}{T_k\left(m_i+m_k\right)}A_n\right) \right)\,.\\
\label{xikvisc}
\end{eqnarray}
\\
Here $A_n=4h_n+\sum_{j+k=n}h_jh_k$ are pure numbers and
\begin{equation*}
B_n(t)\!=\left(A\kappa_{ki}^{\frac{2}{5}}\right)^{\frac{n}{2}}\!\left(2\frac{T_km_i+T_im_k}{m_im_k}\right)^{\frac{n}{20}}\!\left(\frac{\left(20+n\right)n}{800}\right)\!\Gamma\left(\frac{n}{20}\right)\,.
\end{equation*}
In Fig.~\ref{GR2visco}, we show the MSDs for a binary mixture of particles colliding with the time-dependent restitution coefficient. At short times the particles move along ballistic trajectories,  $\left< R^2_k\right>\sim t^2$, and at long times $\left< R^2_k\right>\sim t^{1/6}$. In Fig.~\ref{GR2visco3} the ternary granular mixture is depicted. In both cases, the equipartition of granular temperatures holds at the initial time moment.
Also, for the velocity-dependent restitution coefficient, the agreement between the theory and simulation data is excellent.\\

\begin{thebibliography}{99}

\bibitem{GranRev} H. Jaeger, S. Nagel, and R. Behringer, Granular solids, liquids, and gases. \textit{Rev. Mod. Phys.} \textbf{68}, 1259 (1996).

\bibitem{PhysGranMed} H. Hinrichsen, D.~E. Wolf \textit{The Physics of Granular Media.} (Berlin: Wiley, 2004).

\bibitem{SanPowGr} J. Duran, \textit{Sands, Powders and Grains}. Berlin: Springer-Verlag, (2000).

\bibitem{DryGranMed} H.~J. Herrmann, J.-P. Hovi and S. Luding,  \textit{Physics of Dry  Granular Media.} Dordrecht: NATO ASI Series, Kluwer (1998).

\bibitem{mars} M. P. Almeida, E. J. R. Parteli, J. S. Andrade and H. J. Herrmann. \textit{Proc. Natl. Acad. Sci. USA} \textbf{105}, 6222 (2008).

\bibitem{book} N. V. Brilliantov, T. P\"oschel, \textit{Kinetic theory of Granular Gases}. Oxford: Oxford University Press (2004).

\bibitem{inter} G. Winnewiser, G.C. Pelz \textit{The physics and chemistry of interstellar molecular clouds.} Proceedings of the 2nd Cologne-Zermatt Symposium Held at Zermatt, Switzerland, Springer (1993).

\bibitem{ringbook} R. Greenberg, A. Brahic, \textit{Planetary Rings.} University of Arizona Press, Tucson (1984).

\bibitem{rings} F. G. Bridges, A. Hatzes and D. N. C. Lin, \textit{Nature} \textbf{309}, 333 (1984).

\bibitem{pnas} N.V. Brilliantov, P. L. Krapivsky, A. Bodrova, F. Spahn, H. Hayakawa, V. Stadnichuk, J. Schmidt, \textit{Proc. Natl. Acad. Sci. USA} \textbf{112}, 9536 (2015).

\bibitem{aster} D. Hestroffer, P. Sánchez, L. Staron et al. \textit{Astron. Astrophys. Rev.} \textbf{27}, 6 (2019).

\bibitem{mehta} A. Mehta, in \textit{Granular Physics}, Cambridge University Press (2011); \textit{Granular Matter}, ed. A. Mehta, Springer, Berlin, (2011).

\bibitem{vib1} R. D. Wildman and D. J. Parker, \textit{Phys. Rev. Lett.}, \textbf{88}, 064301 (2002).

\bibitem{vib2} A. Prevost, D. A. Egolf and J. S. Urbach, \textit{Phys. Rev. Lett.}, \textbf{89}, 084301 (2002).

\bibitem{rot} O. Zik, D. Levine, S.G. Lipson, S. Shtrikman and J. Stavans, \textit{Phys. Rev. Lett.}, \textbf{73}, 644 (1994).

\bibitem{el}  I. S. Aranson and J. S. Olafsen, \textit{Phys. Rev. E}, \textbf{66}, 061302 (2002).

\bibitem{magn1}  A. Snezhko, I. S. Aranson and W.-K. Kwok, \textit{Phys. Rev. Lett.}, \textbf{94}, 108002 (2005).  

\bibitem{magn2}  C. C. Maass, N. Isert, G. Maret and C. M. Aegerter, \textit{Phys. Rev. Lett.}, \textbf{100}, 248001 (2008).

\bibitem{drop1} D. Heielmann, J. Blum, H. J. Fraser, and K. Wolling,
 \textit{Icarus} \textbf{206}, 424 (2010).

\bibitem{drop2} P. Born, J. Schmitz, and M. Sperl, \textit{Microgravity} \textbf{3}, 27 (2017).


\bibitem{Sperl} P. Yu, M. Schröter, M. Sperl. \textit{Phys. Rev. Lett.}, \textbf{124} (20), 208007 (2020).

\bibitem{rocket1} K. Harth, U. Kornek, T. Trittel, U. Strachauer, S. Höme, K. Will, and R. Stannarius, \textit{Phys. Rev. Lett.} \textbf{110}, 144102 (2013).

\bibitem{rocket2} E. Falcon, R. Wunenburger, P. Evesque, S. Fauve, C. Chabot, Y. Garrabos, and D. Beysens, \textit{Phys. Rev. Lett.} \textbf{83}, 440 (1999).

\bibitem{rocket3} K. Harth, T. Trittel, S. Wegner, and R. Stannarius, \textit{Phys. Rev. Lett.} \textbf{120}, 214301 (2018).

\bibitem{flights1} M. Noirhomme, A. Cazaubiel, A. Darras, E. Falcon, D. Fischer, Y. Garrabos, C. Lecoutre-Chabot, S. Merminod, E. Opsomer, F. Palencia, J. Schockmel, R. Stannarius, and N. Vandewalle, \textit{Europhys. Lett.} \textbf{123}, 14003 (2018).

\bibitem{flights2} A. Sack, M. Heckel, J. E. Kollmer, F. Zimber, and T. Pöschel, \textit{Phys. Rev. Lett.} \textbf{111}, 018001
(2013).

\bibitem{flights3} S. Tatsumi, Y. Murayama, H. Hayakawa, and M. Sano, \textit{J. Fluid Mech.} \textbf{641}, 521 (2009). 

\bibitem{flights4} E. Falcon, S. Aumaître, P. Evesque, F. Palencia, C. Lecoutre-Chabot, S. Fauve, D. Beysens, and Y. Garrabos, \textit{Europhys. Lett.} \textbf{74}, 830 (2006).

\bibitem{flights5}  M. Leconte, Y. Garrabos, E. Falcon, C. Lecoutre-Chabot, F. Palencia, P. Évesque, and D. Beysens, \textit{J. Stat. Mech.} P07012 (2006).

\bibitem{flights6}  Y. Grasselli, G. Bossis, and R. Morini, \textit{Eur. Phys. J. E} \textbf{38}, 8 (2015).

\bibitem{sat} M. Hou, R. Liu, G. Zhai, Z. Sun, K. Lu, Y. Garrabos, and P. Evesque, \textit{Microgravity Sci. Technol.} \textbf{20}, 73 (2008).

\bibitem{compbook} T. P\"oschel and T. Schwager, \textit{Computational Granular Dynamics}, Springer, Berlin, (2005).

\bibitem{Bird} G. A. Bird, \textit{Molecular Gas Dynamics and the Direct Simulation of Gas Flows, v. 1}. Oxford: Clarendon Press (1994).

\bibitem{rank1} S. A. Matveev, P. L. Krapivsky, A. P. Smirnov, E. E.
Tyrtyshnikov, and N. V. Brilliantov, \textit{Phys. Rev. Lett.} \textbf{119}, 260601 (2017).

\bibitem{rank2} A. Osinsky, J. Comput. Phys. 422, 109764 (2020).

\bibitem{rank3} S. Matveev, N. Ampilogova, V. Stadnichuk, E. Tyrtyshnikov,
A. Smirnov, and N. Brilliantov, Comput. Phys. Commun. 224,
154 (2018).

\bibitem{mediag} A. I. Osinsky and  N. V. Brilliantov, \textit{Phys. Rev. E}, \textbf{105}, 034119 (2022).

\bibitem{ralf} R. Metzler and J. Klafter, \textit{Phys. Rep.} \textbf{339}, 1 (2000).

\bibitem{sok} I. M. Sokolov, \textit{Soft Matter} \textbf{8}, 9043 (2012).

\bibitem{eli} E. Barkai, Y. Garini, and R. Metzler, \textit{Phys. Today} \textbf{65}, 29 (2012).

\bibitem{georges} J.-P. Bouchaud and A. Georges, \textit{Phys. Rep.} \textbf{195}, 127 (1990).

\bibitem{franosh} F. Höfling and T. Franosch, \textit{Rep. Prog. Phys.} \textbf{76}, 046602 (2013).

\bibitem{ido} I. Eliazar and J. Klafter, \textit{J. Phys. A} \textbf{44}, 405006 (2011).

\bibitem{ultraslow} A. S. Bodrova, A. V. Chechkin, A. G. Cherstvy, and R.Metzler, \textit{New J. Phys.} \textbf{17}, 063038 (2015).

\bibitem{jeon} R. Metzler, J.-H. Jeon, A. G. Cherstvy and E. Barkai, \textit{Phys. Chem. Chem. Phys.} \textbf{16}, 24128 (2014).

\bibitem{annapccp} A. Bodrova, A. V. Chechkin, A. G. Cherstvy and R. Metzler \textit{Phys. Chem. Chem. Phys.} \textbf{17}, 21791 (2015).

\bibitem{garzointruder}
E. Abad, S.B. Yuste, V. Garzó, \textit{Granular Matter} \textbf{24}, 111 (2022). 

\bibitem{annaprl} A. S. Bodrova, A. K. Dubey, S. Puri, N. V.  Brilliantov, \textit{Phys. Rev. Lett.} \textbf{109}, 178001 (2012). 

\bibitem{garzoreview} M. García Chamorro, R. Gómez González, V. Garzó. \textit{Entropy}, \textbf{24}, 826 (2022).

\bibitem{garzointruder23} R. Gómez González, E. Abad, S. Bravo Yuste, V. Garzó. \textit{Phys. Rev. E 108}, \textbf{024903} (2023).

\bibitem{garzointruder24} R. Gómez González, S. Bravo Yuste, V. Garzó, arXiv:2409.08726.

\bibitem{anna2024} A. S. Bodrova, \textit{Phys. Rev. E}, \textbf{109}, 024903 (2024).



\bibitem{Saitoh} K. Saitoh, A. Bodrova, H. Hayakawa,  N.V. Brilliantov,  \textit{Phys. Rev. Lett.}, \textbf{105}, 238001 (2010).

\bibitem{lev} A. Bodrova, D. Levchenko, N.V. Brilliantov, \textit{Europhys. Lett.} \textbf{106}, 14001 (2014).

\bibitem{zippelius} H. Uecker, W.T. Kranz, T. Aspelmeier, A. Zippelius, \textit{Phys. Rev. E} \textbf{80}, 041303 (2009).

\bibitem{hrenya} V. Garzo, C.M. Hrenya and J.W. Dufty, \textit{Phys. Rev. E}, \textbf{76}, 031304 (2007).

\bibitem{haff} P. K. Haff, \textit{J. Fluid Mech.}, \textbf{134}, 401 (1983).

\bibitem{brey} J. J. Brey, M. J.  Ruiz-Montero,  R. Garcia-Rojo and J. W. Dufty,  \textit{Phys. Rev. E}, \textbf{60}, 7174-7181 (1999).


\bibitem{m1} S. Burov, J.-H.
Jeon, R. Metzler, and E. Barkai, \textit{Phys. Chem. Chem. Phys.} \textbf{13}, 1800 (2011).

\bibitem{golding} I. Golding and E. C. Cox, \textit{Phys. Rev. Lett.}, \textbf{96}, 092102 (2006)

\bibitem{wang} Y. M. Wang, R. H. Austin, and E. C. Cox, \textit{Phys. Rev. Lett.}, \textbf{97}, 048302
(2006).


\bibitem{sinai} A. Godec, A. V. Chechkin, H. Kantz, E. Barkai, and R. Metzler,
J. Phys. A \textbf{47}, 492002 (2014).

\bibitem{metzlerprl08} Y. He, S. Burov, R. Metzler, and E. Barkai, Phys. Rev.
Lett. \textbf{101}, 058101 (2008); A. Lubelski, I. M. Sokolov, and J. Klafter,
\emph{ibid.} \textbf{100}, 250602 (2008).

\bibitem{schulz} J. H. P. Schulz, E. Barkai, and R. Metzler, Phys. Rev. Lett.
\textbf{110}, 020602 (2013); Phys. Rev. X \textbf{4}, 011028 (2014). 

\bibitem{glass} J.-P. Bouchaud, J. Phys. I France \textbf{2}, 1705 (1992).

\bibitem{mesmollowrank} A. I. Osinsky, \textit{J. Comp. Phys.} \textbf{506 (1)}, 112942 (2024).

\bibitem{nature} N. V. Brilliantov, A. Formella, T. Pöschel, \textit{Nature Communications} \textbf{9 (1)}, 797 (2018).

\bibitem{BirdNanbuCompare} I. D. Boyd, J. P. W. Stark, \textit{Phys. Fluids}, \textbf{30 (12)} 3661 (1987).

\bibitem{rpbs99} R. Ramirez, T. P\"oschel, N.V. Brilliantov, T. Schwager,  \textit{Phys. Rev. E} \textbf{60}, 4465 (1999).

\bibitem{delayed} T. Schwager, T. P\"oschel,  \textit{Phys. Rev. E} \textbf{78}, 051304 (2008).

\bibitem{bshp96} N.V. Brilliantov, F. Spahn, J.M. Hertzsch, T. P\"oschel,  \textit{Phys. Rev. E} \textbf{53}, 5382 (1996).

\bibitem{goldobin} D. S. Goldobin, E. A. Susloparov, A. V. Pimenova, N. V. Brilliantov, \textit{Eur. Phys. J. E} \textbf{38}, 55 (2015).

\bibitem{Os} A.S. Bodrova A. Osinsky, N.V. Brilliantov, \textit{Sci. Rep.} \textbf{10}, 693  (2020).



\end{thebibliography}
\end{document}